\documentclass[12pt]{article}
\usepackage{epsfig, amssymb, graphicx, amsmath} 
\usepackage{amsthm}
\usepackage{relsize}
\usepackage{url}
\usepackage{comment}
\usepackage[round-mode=places, round-integer-to-decimal, round-precision=2,
table-format = 1.2, 
table-number-alignment=center,
round-integer-to-decimal,
output-decimal-marker={,}
]{siunitx} 
\usepackage{booktabs}
\usepackage{enumitem}
\usepackage{eurosym}	
\usepackage{wrapfig}	
\usepackage{float}		
\usepackage{subfig}	
\usepackage{footnote}
\usepackage{thmtools}
\usepackage{bbm}		
\usepackage[title]{appendix}
\usepackage{adjustbox}
\usepackage{optidef}
\usepackage{xcolor,colortbl}
\usepackage{algorithm}
\usepackage{algpseudocode}
\usepackage{multirow}
\usepackage[authoryear]{natbib}
\bibpunct{(}{)}{,}{a}{}{,}

\usepackage{amsmath}

\DeclareMathOperator*{\argmin}{arg\,min}

\theoremstyle{definition}

\theoremstyle{remark}

\numberwithin{theorem}{section}
\numberwithin{proposition}{section}
\numberwithin{lemma}{section}
\numberwithin{corollary}{section}
\numberwithin{definition}{section}
\numberwithin{remark}{section}
\numberwithin{example}{section}

\newcommand{\be}{\begin{equation}}
\newcommand{\en}{\end{equation}}
\newcommand{\ben}{\begin{equation*}}
\newcommand{\enn}{\end{equation*}}
\newcommand{\bea}{\begin{eqnarray}}
\newcommand{\ena}{\end{eqnarray}}

\begin{document}
	
	\newlength\tindent
	\setlength{\tindent}{\parindent}
	\setlength{\parindent}{0pt}
	\renewcommand{\indent}{\hspace*{\tindent}}
	
	\begin{savenotes}
		\title{
			\bf{ 
 Physical Climate Risk in Asset Management				
		}}
		\author{
			Michele Azzone$^\ddagger$ \& 
			Matteo Ghesini$^\ddagger$ \&
                Davide Stocco$^+$ \& 
						Lorenzo Viola$^*$ 
		}
		
		\maketitle
		
		\vspace*{0.11truein}
		\begin{tabular}{ll}
			$(\ddagger)$ &  Politecnico di Milano, Department of Mathematics, 32 p.zza L. da Vinci, Milano \\
            $(+)$& CREST, ENSAE, Institut Polytechnique de Paris, 5 Avenue Henry Le Chatelier, Palaiseau\\
   		$(*)$ & Arca Fondi SGR, 3, Via Disciplini, Milano \\
		\end{tabular}
	\end{savenotes}
	
	\vspace*{0.11truein}
	\begin{abstract}
		\noindent
Climate-related phenomena are increasingly affecting regions worldwide, manifesting as floods, water scarcity, and heat waves, significantly impairing companies’ assets and productivity. It is essential for asset managers to quantify the exposure of their portfolios to such risk. To this aim, we develop a framework, based on the Vasicek model for credit risk, that introduces downward jumps due to climate phenomena in a company asset’s dynamics. These negative shocks are designed to mirror the negative effect of extreme climate events.  The model calibration  relies  on companies' asset intensity and geographical exposure. We apply the new multivariate firm value model with jumps to assess the impact of climate-related extreme events on expected and unexpected portfolio losses. Our findings indicate that expected losses increase over time, with pronounced differences in exposure observed across sectoral indices. From an environmental policy perspective, these results suggest the need for additional capital buffers to offset losses arising from physical climate risks, particularly in sectors with high asset intensity.
	\end{abstract}
	
	\vspace*{0.11truein}
	
	\vspace*{0.04truein}
	
	\vspace*{1cm}
	\begin{center}
		\Large\bfseries 
Running title: Physical Climate Risk in Asset Management				
	\end{center}
	\vspace{1cm}
	\begin{flushleft}
		{\bf Address for correspondence:}\\
		Michele Azzone\\
		Department of Mathematics \\
		Politecnico di Milano\\
		32 p.zza Leonardo da Vinci \\ 
		I-20133 Milano, Italy \\
		michele.azzone@polimi.it
	\end{flushleft}
{\small The views expressed are those of the author and do not necessarily reflect those of Arca Fondi SGR.}
	
	\newpage

	\vspace*{0.1truein}

\section{Introduction}
\label{}


Physical risk emerging from climate change will play a key role in the asset management and banking industry. 
It comprises both acute risks stemming from climate-related hazards (e.g., from extreme weather events) and chronic risks  (e.g., from sea-level rise) from long-run climate impacts \citep{chou2023firms,de2024we,bressan2024asset}. 
In response, central banks and supervisory authorities globally expect financial institutions with material climate-related and environmental exposures to integrate physical climate risks into their risk management frameworks.\footnote{See e.g., the ECB \url{https://www.ecb.europa.eu/press/economic-bulletin/focus/2023/html/ecb.ebbox202302_06~0e721fa2e8.en.html}, the FED \url{https://www.federalreserve.gov/publications/files/csa-exercise-summary-20240509.pdf}, and the Basel Committee on Banking Supervision perspectives  \url{https://www.bis.org/bcbs/publ/d560.pdf}.}
We propose a new and simple model to include this risk in asset management decisions, using a multivariate approach. 
We model physical risk impact on stock portfolios, aggregating the firm-level effects of such risk. 
To the best of our knowledge, this is the first model in the literature that exhibits these characteristics and that can be flexibly applied to any stocks portfolio using widely available data.


 
Physical risk affects directly firms' tangible assets (e.g. production plants or equipment) as well as operating costs, such as relocating and insurance costs.
These shocks result in lower profits and reduced repayment capacity, which are then transmitted to the listed stock prices \citep{ginglinger2023climate}. 
To mirror this cascading effect we base our approach on   \citet{vasicek1987probability}, a multivariate generalization of the simple \citet{Merton74} Firm Value Model (FVM) that considers a single market risk factor. 
In this framework, the firms' asset value is modeled with a geometric Brownian motion, while equity is modeled as a European call option on the firm's asset value, with the company's debt as strike price.
We propose a modification of the \citet{vasicek1987probability} model that incorporates jumps to capture the impact of climate events which adversely affect a firm's asset value.
The jump process frequency and size depend on the firm's characteristics such as its geographic exposure and how much of its revenues are linked to physical assets.


In our study, we follow an efficient and versatile approach for the calibration. It is based on widely available data that covers most listed stocks worldwide and enables low-cost implementation by financial institutions.
Specifically, we focus on the exposure of physical assets to climate risks. 
We utilize the publicly available ND-GAIN (Notre Dame Global Adaptation Initiative) Country Index  to assess risk based on geographical location and incorporate balance sheet data to estimate the proportion of a company’s assets exposed to such risks.\footnote{See \url{https://gain.nd.edu/assets/522870/nd_gain_countryindextechreport_2023_01.pdf}.} The ND-GAIN index is a widely used benchmark for climate risk in the literature \citep[see e.g.,][]{russo2019half, dechezlepretre2022fighting}. 
However, our model is very flexible since it can be calibrated starting from different definitions of climate losses such as the ones computed by \citet{bressan2024asset} for Mexican companies or from private data providers such as those listed in \citet{hain2022let} or the CRIS methodology discussed in \citet{ginglinger2023climate}.

Once the extended \citet{vasicek1987probability} model is calibrated, we can utilize it to compute different climate-related risk metrics for a portfolio. In our analysis, we focus on computing an add-on for the Value-at-Risk (VaR) of a portfolio that accounts for a climate-stressed scenario. We then evaluate the evolution of such metrics for different time horizons. This is particularly relevant from the perspective of financial institutions, as it enables the incorporation of physical risks into their risk management practices building upon their existing frameworks. Moreover, from an environmental policy perspective, these findings suggest that regulators and supervisors should introduce additional capital requirements for exposed portfolios to offset such risks.

\subsection*{Climate Change Implications}
The  Intergovernmental Panel on Climate Change (IPCC)  Synthesis Report confirms that more than a century of fossil fuel combustion, along with unsustainable and unequal energy and land use, has unequivocally caused global warming. Between 2011 and 2020, the global surface temperature reached 1.1°C above 1850–1900 levels, resulting in widespread adverse impacts and losses for both people and ecosystems. The decade from 2015 to 2024 was the warmest on record, with 2024 being the warmest year and 2016 being the second warmest.
Six global temperature datasets show the five-year average (2020-2024) as the highest on record. The ten warmest years in the 175-year record have all occurred within this decade. 
Current nationally determined contributions imply that global warming will reach 1.5°C in the early 2030s, significantly reducing the likelihood of limiting temperature rise to 2.0°C by the end of the century. Each additional increment of warming will intensify multiple and overlapping climate-related hazards across all world regions.

Consequently, the last decade saw a sharp increase in extreme weather events, such as heatwaves, floods, and wildfires, leading to unprecedented economic losses, with several disasters exceeding USD 10 billion. Ice sheet loss in the Arctic and Antarctic accelerated, contributing to rising sea levels. CO2 concentrations reached record highs, further intensifying climate change. 
In recent years, wildfire activity across the globe has reached unprecedented levels, causing record-breaking burn areas and carbon emissions, particularly in regions such as Canada, Amazonia, and the Mediterranean, where extreme events were estimated to be up to 20 times more likely than in a pre-industrial climate \citep{jones2024state}.
The impacts of climate change have disproportionately affected the most vulnerable people and systems, particularly in least developed countries in the Global South, despite their low per capita emissions. Vulnerability is highest in regions with poverty, limited services, weak governance, conflict, and climate-sensitive livelihoods, such as parts of Africa, South Asia and Central and South America. Human and ecosystem vulnerabilities are deeply interconnected and shaped by inequity, unsustainable practices, and historical injustices \citep{damert2018external}.

Despite improved early warning systems, climate-related displacement surged, highlighting the growing risks and the urgent need for climate adaptation and mitigation strategies. 
Numerous initiatives have been introduced by various countries and international organizations - such as the United Nations Framework Convention on Climate Change, the United Nations Environment Programme, and the IPCC - to address global climate change through both mitigation and adaptation strategies. 
A mitigation strategy refers to actions taken to reduce or prevent the emission of greenhouse gases into the atmosphere, including activities such as carbon sequestration, the Clean Development Mechanism, joint implementation, as well as  the promotion of renewable and non-polluting energy sources such as solar, wind, and geothermal power \citep{wang2023climate}.
On the other hand, adaptation strategies include the processes of adjustment to actual or expected climate changes and its effects, in order to moderate harm or exploit beneficial opportunities.\footnote{For a comprehensive analysis of climate change mitigation and adaptation strategies, we refer the reader to the IPCC Sixth Assessment Report (2022), available at: \url{https://www.ipcc.ch/report/ar6/wg2/downloads/report/IPCC_AR6_WGII_SummaryVolume.pdf}}
These adverse effects can have severe financial consequences for firms that can be adversely affected if ill-prepared  \citep{gasbarro2016corporate}.

\subsection*{Risk and Climate Risk}
Risk, in general, refers to the uncertainty surrounding future outcomes, encompassing both the possibility of loss and gain. In financial contexts, risk specifically denotes the potential for monetary loss due to adverse movements in financial variables such as interest rates, exchange rates, or asset prices  \citep{mcneil2015quantitative}. Financial risk is multifaceted and includes market risk, credit risk, liquidity risk, and operational risk, each of which can significantly affect the stability and performance of firms and financial systems. Understanding and quantifying these risks is central to financial decision-making and regulatory frameworks.  As noted by \citet{jorion2007good}, effective risk management requires a systematic approach to identifying, measuring, and mitigating exposures in an increasingly complex and interconnected financial environment.

A new class of risk that spans all aforementioned traditional  categories of financial risk is {climate risk}. Climate risk is commonly categorized into two main components: {physical risk} and {transition risk}.
Physical risk refers to the potential economic and financial losses that arise from the physical impacts of climate change events. These risks stem from both acute events (such as hurricanes, floods, wildfires, and heatwaves) and chronic changes (such as rising sea levels, long-term temperature increases, and shifts in precipitation patterns).
The first class represents short-in-time but severe events, while chronic changes are identified by longer-term shifts in climate patterns.  
We can define climate transition risk as the unanticipated financial gains or losses arising from a faster-than-expected transition toward a carbon-neutral economy.
 It includes risks created by mitigation and adaptation policies, emerging clean technologies, and behavioral changes of consumers and investors \citep{bolton2023global,fliegel2025not}: some firms will benefit from the shift to a low-carbon economy, while others, particularly those heavily reliant on fossil fuel-based activities, will incur significant losses.

In this paper, we focus on the physical climate market risk, i.e. how extreme weather phenomena could affect a portfolio of stocks.
Physical climate risks affect asset values through several key transmission channels. Acute physical events, such as floods, hurricanes, or wildfires, can lead to the destruction of physical assets, disruption of operations and supply chain, and increased insurance costs. Chronic risks, including rising temperatures and sea-level rise, can gradually reduce land usability, agricultural productivity, and labor efficiency. These physical impacts can translate into lower revenues, higher operating costs, or even stranded assets for affected firms. In turn, this deteriorates firms' financial performance and creditworthiness, leading to downward revisions in equity valuations and increased default risk on debt instruments \citep[see][and references therein]{bressan2024asset}. At the systemic level, widespread physical damages can trigger portfolio reallocation, risk repricing, and higher market volatility, especially in sectors or regions with high exposure to climate hazards.

\subsection*{Asset Management in a Climate Change Scenario}
The aim of this work is to develop a reliable model for assessing physical climate  market risk within the context of asset management. In particular, we focus on asset managers, broadly defined as financial institutions or professional entities that manage investments on behalf of clients—including individuals, pension funds, insurance companies, and sovereign wealth funds—with the goal of optimizing returns for a given level of risk. Within this framework, we consider financial firms that invest in diversified portfolios of equities. The objective is to quantify how physical risks stemming from climate change - such as extreme weather events, temperature anomalies, or rising sea levels - can propagate through market channels, impacting the value and risk profile of portfolios under management.

The majority of the literature on climate risk impact on portfolios focuses on transition risk and not on physical risk.
\citet{le2024corporate} utilize a FVM that accounts for the uncertainty of the carbon tax trajectory while \citet{kolbel2024ask}  show that, in a FVM approach, CDS spread are affected by transition and physical risk news.
\citet{de2023climate} discuss the influence of green investors on long-run emissions trajectories, finding that  
tighter climate regulations and climate-related technological innovations abate carbon emissions,  in contrast to  uncertainty regarding future climate.
\citet{bolton2023global} point out that transition risk is not significantly related to different exposures to physical risk.

To the best of our knowledge, only two contributions specifically address the impact of physical risk on financial assets. \citet{dietz2016climate} estimate a global VaR for financial assets by employing a single dividend-discounted cash flow model, where the growth rate is adjusted according to GDP shocks derived from the DICE2010 model. However, we note that this global VaR may be overly simplistic to apply in practice, as it does not differentiate between the various risks across different asset classes and sectors.

\citet{bressan2024asset} propose an asset-level assessment of climate losses, which could potentially address the issue by highlighting assets with differing levels of exposure. Their approach aims to accurately model the impact of extreme climate events on a sample of companies based in a single country (Mexico), where point-specific physical exposures can be considered (e.g.,  a company's plant located near a river is exposed to severe flooding risk), and integrates this estimation into a dividend-discounted cash flow model. However, from a practitioner's perspective, this model may be challenging to implement in practice, as estimating the losses for a diversified portfolio of assets requires detailed, granular, and costly data. Furthermore, given the significant difficulty in predicting climate evolution and its effects, along with the uncertainty inherent in such forecasts, practitioners may question whether such a detailed modeling approach justifies the financial and time investment required.


\subsection*{Main contributions}
This paper adds to the existing literature by
proposing an alternative approach for physical risk in asset management that strikes a balance between the two extrema discussed above: \citet{dietz2016climate} estimate only a global loss, while \citet{bressan2024asset} provide highly detailed asset-level estimations for a single country, which are challenging to apply for portfolio risk assessment. Our model, in contrast, can be easily applied to any portfolio of assets for which climate loss estimates and balance sheet data are available, providing various climate-related risk metrics at both the portfolio and asset levels. We also outline a framework for estimating the aforementioned climate losses using the publicly available ND-GAIN dataset, though we emphasize that our approach can be easily adapted to consider other definition of climate losses, such as the ones of \cite{bressan2024asset}. Furthermore, our  model can be extended to include more detailed considerations at corporate level, such as assessing collateral deterioration or other channels of physical risks, particularly relevant for sectors like banking.

Overall, this paper makes three main contributions. First, we introduce a novel firm value-based modeling framework for estimating physical risk at the firm level, which can be easily integrated into financial institutions' decision-making processes and customized for various climate risk metrics. Second, we propose a calibration procedure for this model, leveraging estimated climate risk losses and widely available balance sheet data. Third, we demonstrate that the model can be used to compute different financial risk metrics at the portfolio level, such as an add-on to the portfolio VaR that captures physical climate risk. These contributions not only fill a gap in the existing literature but also address the growing need for financial institutions, driven by regulatory pressures, to incorporate physical risks into their risk management frameworks.


\section{The firm value model}
\label{Model}

We use a multivariate FVM to relate the asset value and the equity value of companies in a \citet{vasicek1987probability} framework. 
There are several applications of \textit{univariate} FVM with jumps in the literature starting from the seminal works of 
\citet{zhou2001term,zhang2009explaining}.
The contributions closest to ours are  \citet{le2024corporate}, which apply a FVM with jumps to corporate bonds in the case of transition risk, and \citet{kolbel2024ask}, which estimate the impact of regulatory climate risk disclosures (both physical and transitional) on credit default swap market.
Differently from these papers, we consider a more realistic \textit{multivariate} framework and we calibrate the model directly under the martingale measure $\mathbb{Q}$ as it is standard in the option pricing literature \citep[see e.g.,][]{hull2016options}. 
This allows us to match the estimated climate losses without mixing historical and martingale measures.

We denote with $J$  the set of companies in the market. 
For a given company $j \in J$, we call $T_j$ the average duration of its debt. Consistently with the \cite{vasicek1987probability} framework we model the equity value  $E_{j}(T_j)$ at $T_j$ as
\begin{equation}
	\label{eq:DefaultCondition}
	E_{j} (T_j) = 
	\begin{cases}
		V_{j}(T_j) - D_j & \text{if $V_{j} (T_j) > D_j$} \\
		0 & \text{otherwise}
	\end{cases}\;\;,
\end{equation}
where $V_{j}(T_j)$ is the total asset value,  and $D_j$ is the total debt of the corporate $j$. Notice that  $E_{j} (T_j) $ definition is equivalent to the payoff of a European call on $V_{j} (T_j)$ with strike $D_j$. Its value at time $t<T_j$ depends on the dynamic of $V_{j} (T_j)$, which follows a geometric Brownian motion described by the stochastic differential equation
\begin{equation*}
\hspace{-1.4cm}
d\left(\log \frac{V_{j} (t)}{V_{j} (0)}\right) =
\left(r_j - \frac{\sigma^2_j}{2} - \frac{\omega_j^2}{2}\right)dt+\sigma_jdW_{j}(t) + \omega_jdZ(t)  \nonumber
\end{equation*}

where $r_j$ is the risk-free rate of the country in which the company operates, and $\sigma_j$ and $\omega_j$ are the volatility terms of two independent standard Brownian motions $W_{j}(t)$ and $Z(t)$. $W_{j}(t)$  captures idiosyncratic firm-specific risk, while $Z(t)$ represents a systematic market-wide risk component.
The solution of the above SDE is
\begin{equation}
	V_{j} (t) = V_{j}(0) e^{\left(r_j - \frac{\sigma^2_j}{2} - \frac{\omega_j^2}{2}\right)t + \sigma_jW_{j}(t) + \omega_jZ(t) }\;\;.\label{eq:MultiMertonSDE_Sol}
\end{equation}
To ensure that the correlation between the logarithm of asset values of any two distinct firms is $\rho\in [0,1]$, we select $\sigma_j$ and $\omega_j$ such that 
$\text{cor}\left(\log {V_i(t)},\log {V_j(t)}\right) = \rho $, $\forall i \neq j$. The $\sigma_j$ and $\omega_j$  that satisfy this condition are 
\begin{equation}
	\label{eq:Rel_Chap2}
	\begin{aligned}
		\omega_j   &=: \hat{\sigma}_j \sqrt{\rho}  \\
		\sigma_j &=: \hat{\sigma}_j \sqrt{1-\rho}
	\end{aligned}\;\;,
\end{equation}

where $\hat{\sigma}_j $ is   a positive constant.
This simple choice for describing market correlations aligns with \citet{gordy2000comparative, gordy2003risk}, a benchmark for risk management models worldwide.\footnote{This model is the reference for the VaR formula in the Basel methodology, see \url{https://www.bis.org/publ/bcbs_wp22.pdf}.} Considering the dynamic of $V_j(t)$ in equations \eqref{eq:MultiMertonSDE_Sol} and \eqref{eq:Rel_Chap2} the risk-neutral value of the equity at time $t<T_j$ is 
\begin{align}
    E_j(t)&=e^{-r_j(T_j-t)}\mathbb{E}_t\left[(V_j(T_j)-D_j)^+\right]\nonumber\\&=C_{BS}(V_j(0),D_j,\hat{\sigma}_j,T_j-t,r_j)\;\;,\label
    {eq:BSFormula}
\end{align}
where $C_{BS}(\bar{S},\bar{K},\bar{\sigma},\bar{T},\bar{r})$ is the Black and Scholes (BS)
 European call  price with spot price $\bar{S}$, strike price $\bar{K}$, maturity $\bar{T}$ and risk-free rate $\bar{r}$. The BS formula depends on the volatility $\hat{\sigma}_j$ because the exponent of $V_j(t)$ is Gaussian with variance $\hat{\sigma}_j^2 t$.

In the following, we present the model for the firms' asset value when we consider a climate-stressed scenario. We select   two key drivers to estimate the potential exposure to physical risk following the assessment of the ECB   for European countries: the locations of firms’ physical assets and how vulnerable are the assets of a firm to natural hazards.\footnote{\url{https://www.ecb.europa.eu/stats/all-key-statistics/horizontal-indicators/sustainability-indicators/data/html/ecb.climate_indicators_physical_risks.en.html#potential_exposure}}

Under these assumptions, firms that are located in regions with similar exposures and share similar asset structures are subject to comparable physical risks stemming from climate change \citep{bressan2024asset}. The natural metric that we utilize for quantifying the asset structure of a firm is what we call its asset intensity. 
Asset intensity is a measure of how much a company relies on its physical assets to generate revenues. 
Similarly to \citet{bressan2024asset}, we define it as the ratio between the tangible assets of a company (in particular we utilize the property, plant, and equipment entry from the balance sheet) and its revenues.
We use this ratio since the total assets of a company would include various intangible assets that are not directly influenced by physical climate risks. Firms with the lowest asset intensity are typically found in the funds and IT sectors, whereas mining and heavy industry companies exhibit the highest asset intensity. It is important to note that this metric may not fully capture indirect exposure to climate risks, particularly for firms in the banking and insurance sectors. A refined classification approach, such as the establishment of distinct clusters for financial institutions, could enhance the accuracy of the method.

We include these two drivers by dividing the firms into clusters with similar climate risk exposures. Each firm $j$ $\in$ $J$ belongs to a climate cluster $k$ $\in$ $K$ which contains all companies with similar asset intensity and  similar geographical exposures.  We model the  dynamic of the climate-stressed asset value $\tilde{V}_{j}(t)$ of firm $j$ in cluster $k$ as
\begin{equation}
	\label{eq:JumpMultiMertonSDE_Sol}
	\tilde{V}_{j}(t)= V_{j}(0) e^{ \left(r_j - \frac{\sigma^2_j}{2} - \frac{\omega_j^2}{2}\right)t + \sigma_jW_{jt} + \omega_jZ_t + L_{k}(t)}\;\;,
\end{equation}
where  $L_k(t)$ is a cluster-specific Compound Poisson process with only negative jumps that represents the physical climate  losses on the firm's assets in the climate-stressed scenario.
The Compound Poisson process $L_k(t)$ of  cluster $k$ is defined as

\begin{equation}
	\label{eq:Compund_2}
	L_k(t)=  \sum_{n=1}^{N_k(t)}Y_{kn}\;\;,
\end{equation} 

where $N_k(t)$ is a Poisson process with intensity $\lambda_k$ describing the frequency of jumps, and $\{Y_{kn}\}_n$ is a sequence of negative random variables describing the amplitude of the downward jumps. We consider the simplest case of jumps with constant (negative) amplitude $\theta_k$.\footnote{Exponential and Gamma distributions have also been considered for numerical experiments providing similar results.} We emphasize that the Compound Poisson has only negative jumps, as we aim to capture the asset values's exposure towards adverse effects  of climate change  that are not yet reflected in market prices.
As discussed in the introduction, the advantage of using a FVM approach is the possibility of considering climate-related shocks on the asset value, which is a key channel of transmission of physical risk.



To evaluate the equity value of company $j$ in the climate-stressed scenario we cannot use the Black-Scholes formula as in \eqref{eq:BSFormula} to price the call option since the firm's value is a jump-diffusion process. Instead, we utilize the \cite{lewis2001simple} formula. 
Lewis’ approach (based on Fourier transform methods) allows pricing in general settings, including stochastic volatility and jump processes, in particular when the characteristic function of the log-process is known. While Black–Scholes yields simple analytical formulas under strong assumptions, Lewis provides a flexible framework to derive option prices semi-analytically from the characteristic function of the underlying process. From a computational point of view, the Lewis formula is comparable to the Black and Scholes one because it involves just an integration in the complex plane.
The equity value, in the climate-stressed scenario is
\begin{small}
\begin{align}
    \Tilde{E}_j(t)&=e^{-r_j(T_j-t)}\mathbb{E}_t\left[(\tilde{V}_j(T_j)-D_j)^+\right]\nonumber\\&=C_{LW}(V_j(0),D_j,\hat{\sigma}_j,T_j-t,r_j,\theta_k,\lambda_k)\nonumber\\
   &= V_j(0) e^{-\gamma_k T} \left[ 1 - \frac{e^{-\frac{z_j}{2}}}{2\pi}\int_{-\infty}^{+\infty}  \frac{e^{-i v z_j} \Phi(-v - \frac{i}{2})}{v^2 + \frac{1}{4} } dv\right]\label{eq:option price}
\end{align}
\end{small}

where $\Phi(v)$ is the characteristic function of the logarithm of the asset value in \eqref{eq:JumpMultiMertonSDE_Sol} given by
\begin{equation*}
\Phi(v) = e^{ i v \left(\gamma_j-\frac{1}{2}\hat{\sigma}^2_j\right) T - \frac{1}{2} \hat{\sigma}^2_j v^2 T - \lambda_k\left(1 - e^{-i v \theta_k}\right)T}
\end{equation*}

$z_j = \log \frac{V_j(0)}{D_j} +(r_j-\gamma_k)T$ and $\gamma_k =\lambda_k(1-e^{-\theta_k})$.
 

When analyzing portfolio losses in a climate-stressed scenario we consider the following simulation framework.
We simulate the jump-diffusion process in \eqref{eq:JumpMultiMertonSDE_Sol} on a discrete time grid. We simulate the increments of the Gaussian market factor $Z(t)$ and the idiosyncratic factor $W_j(t)$. For the jump component at the cluster level, $L_k(t)$, we simulate the increments of the Poisson process. 
The detailed description of the simulation algorithm for the jump-diffusion process (with and without jumps) is available in Algorithm \ref{alg:Simulation_Jump_Asset}. In our numerical analysis, we consider $10^5$ simulations and a monthly time-step.

\section{Dataset Composition}
\label{Dataset}

The data set comprises financial and climate data for a large number of companies distributed all over the world.
Company-level data are provided by LSEG; the dataset has been downloaded in  December 2023.
We cover the main MSCI indexes, which encompass a broad spectrum of regions and market segments.\footnote{We consider aggregating indexes such as those for Europe, Nordic countries, Africa, the US Investible Market Index (IMI), the World Index, Emerging Markets, World IMI Small Cap, and Emerging Markets IMI Small Cap. Furthermore, we consider some additional indexes for Argentina, Australia, Brazil, Canada, Chile, China, Hong Kong, Japan, Korea, Kuwait, Malaysia, Mexico, Qatar, Saudi Arabia, Singapore, South Africa, Switzerland, and the United Kingdom. }
Our final dataset comprises 10,430 distinct listed companies  globally. 
Prices and accounting data in the national currency have been converted to USD.

On the financial side, we consider  equity value,  asset value and returns for each stock as well as  total debt, debt's average maturity and EBITDA for estimating the rate of growth of the corporate.
We use these quantities for estimating the FVM and the climate losses.
The risk-free rate is related to the currency associated to companies and missing values are treated considering the average risk-free rate of all companies belonging to the same index. 

On the climate side, geographical location and asset intensity are used to group companies into climate clusters. We consider the geographical location of the listed companies' headquarters and its operational sector (NACE sector). 
We are aware that considering the headquarters of the company as the location of its assets is a limitation from a climate-risk perspective; however, our framework can be easily extended to consider asset level impact of extreme climate events if more granular data are available.
In the results discussed in this paper, we limit ourselves to this simpler assumption that can be easily implemented with the data available on LSEG.
As already mentioned, we consider the ratio of Property, Plant, and Equipment (PPE) over revenues as a measure of asset intensity.

To evaluate the physical risk in different geographical areas, we consider the vulnerability score component provided by the Notre-Dame Global Adaptation Initiative (ND-GAIN).\footnote{For further detail refer to: \url{https://gain.nd.edu/our-work/country-index/}. We utilize the dataset available on December 2023.}
The vulnerability score reports a country's exposure, sensitivity, and capacity to adapt to the negative effects of climate change. ND-GAIN aims to quantify the overall vulnerability by considering six life-supporting sectors – food, water, health, ecosystem service, human habitat, and infrastructure. 
We report the dataset descriptive statistics in Appendix A.

\begin{figure}[t]
	\centering
	\includegraphics[scale=0.7]{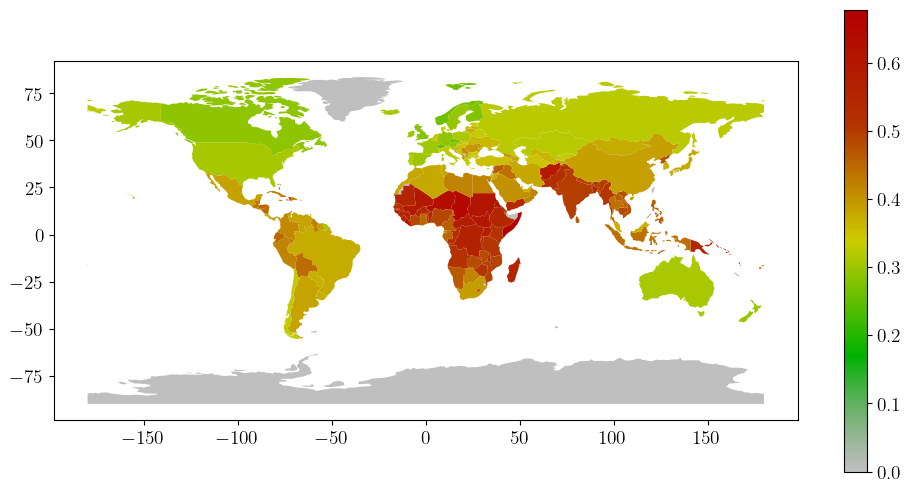}
	\caption[Global Distribution of Vulnerability Scores]{\raggedright A representation of vulnerability scores across countries worldwide, reflecting their susceptibility to climate-related risks. Darker shades indicate higher vulnerability. Countries in grey lack available data.}
	\label{fig:Vulnerability_WorldMap}
\end{figure}

In Figure~\ref{fig:Vulnerability_WorldMap}, we show the global distribution of the vulnerability score.  Red (green) shades indicate higher (lower) vulnerability. Regions around equatorial line, such as central Africa and South Asia, exhibit heightened risk. Such regions are the most affected by climate phenomena and, at the same time, are among the poorest ones, therefore lacking sufficient financial resources to manage these changes.\footnote{See {https://www.imf.org/en/Blogs/Articles/2022/03/23/blog032322-poor-and-vulnerable-countris-need-support-to-adapt-to-climate-change}} These combined factors result in generally high vulnerability scores. 
Interestingly, the most exposed countries are those part of the Global South, while the least vulnerable are part of the Global North.\footnote{For the definition of the Global South we refer to the document of the United Nations, Geneva, 2018: \url{https://unctad.org/system/files/official-document/osg2018d1_en.pdf}}

\section{Calibration}
\label{Cluster_Shock}

In this section, we identify the climate clusters and discuss the methodology we propose for computing average climate losses in each cluster. Then, we calibrate the stressed FVM on such clusters and losses. 
\subsection{Climate Clustering}
\label{Cluster_vul}

To cluster firms from a physical risk perspective, we consider two key drivers: the vulnerability of the firm's country to extreme climate change events and the firm's sector average asset intensity.

For country vulnerability, we rely on the vulnerability component of the ND-GAIN index, which quantifies the average physical risks of climate change faced by countries. This allows us to group countries with similar risk profiles.


We employ a hierarchical clustering methodology \citep[cf.][ch.8, pp.221-239]{Nielsen06}, which results in the identification of three distinct clusters, namely Low, Medium, and High vulnerability. The High vulnerability cluster comprises only a small number of countries located in Central Africa, where a minimal proportion of the companies represented in our dataset is based.
Therefore,
we decide to merge the Medium and High clusters. We call this new cluster Mid-High. 

In Table \ref{table:Vulnerability_Table_Cluster_2}, we report the number of available companies, the average vulnerability score, and the standard deviation for the Low and Mid-High clusters.
We notice that both clusters contain around 4000 firms and that the average vulnerability of the two clusters is separated by more than two standard deviations. 

\begin{figure}[t]
	\centering
	\includegraphics[scale=0.5]{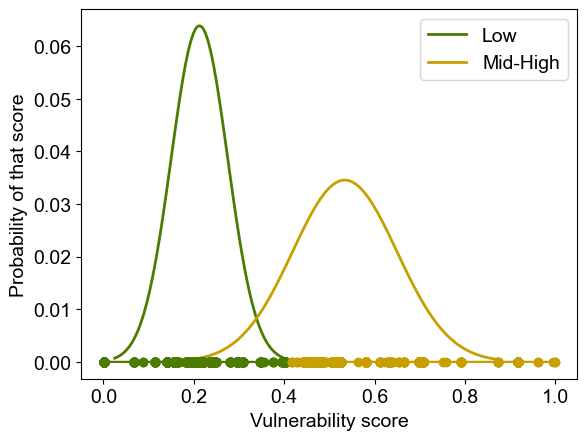}
	\caption[Distribution of clusters]{\raggedright Estimated distributions of the vulnerability scores in the Low (green line) and Mid-High (orange line) clusters. On the bottom,  we report each observation of the vulnerability scores (dots).}
	\label{fig:Distribution_of_clusters}
\end{figure}

\begin{table}[t]
	\scriptsize
	\centering 
	\begin{tabular}{l c c c}
		\toprule
		& \textbf{Available Companies} & \textbf{Average Vulnerability}&\textbf{Standard Deviation}\\
		\midrule
		\rowcolor{gray!15}
		Low & 4599 & 0.212 & 0.062 \\  
		Mid-High & 3824 & 0.534 & 0.115 \\
		\bottomrule
	\end{tabular}
	\\[10pt]
	\caption[Descriptive Analysis of Vulnerability Clusters - 2 Clusters]{\raggedright Descriptive analysis of the vulnerability score for the  Low and Mid-High risk clusters. We report the number of available companies, the average vulnerability score, and the standard deviation.}
	\label{table:Vulnerability_Table_Cluster_2}
\end{table}
In Figure \ref{fig:Distribution_of_clusters}, we report the estimated distribution of the two clusters' vulnerability scores. We point out that the Low risk cluster is concentrated around the average score of 0.2 while the Mid-High risk one is more dispersed.

\begin{figure}[t]
	\centering
	\includegraphics[scale=0.5]{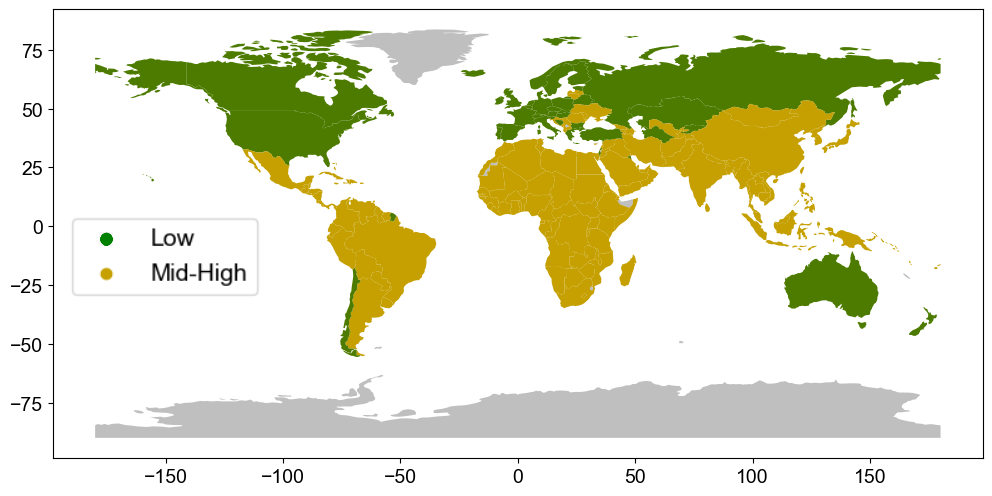}
	\caption[Clusterization of Vulnerability Scores - 2 Clusters]{\raggedright Vulnerability score clustering on the world map: Low Risk (Green) and Mid-High Risk (Orange). Countries in grey lack available data.}
	\label{fig:Vulnerability_WorldMap_Cluster_2}
\end{figure}

In Figure \ref{fig:Vulnerability_WorldMap_Cluster_2}, we show the geographical distribution of the two clusters.
We point out that the Mid-High risk cluster  vulnerability score corresponds almost perfectly to the countries belonging to the Global South, while the other one to the Global North.
This reflects the different impacts of climate change across different countries.

To measure asset intensity, we first classify firms according to their NACE Rev. 2 sectors (at the two-digit granularity level) and calculate sector-level average asset intensity. Subsequently, we perform cluster analysis on these sector-level averages to identify distinct groupings based on asset intensity. Prior to clustering, we preprocess the data to mitigate the influence of outliers by winsorizing extreme values (at the 1st and 99th percentiles) and we normalize in the [0, 1] interval using min-max scaling.

In Table \ref{table:Intensity_Table_Cluster}, we report the number of companies, the average asset intensity, and its standard deviation (at the NACE sector level) for each cluster. We notice how the Medium group is the mostly populated with an average intensity of 0.071 and that Low and Extreme cluster are similarly populated, with an intensity of 0.032 and 0.345, respectively.

\begin{figure}[H]
	\centering
	\includegraphics[scale=0.5]{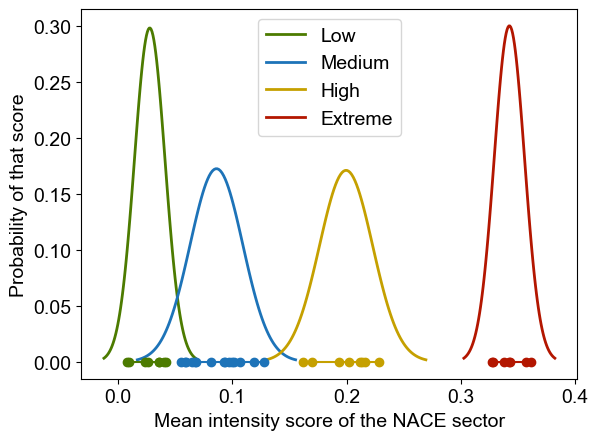}
	\caption[Distribution of clusters for intensity]{\raggedright Estimated distributions of the asset intensity in the Low (green line), Medium (blue line), High (orange line), and Extreme (red line) clusters. On the bottom,  we report each observation of the asset intensity (dots) at the NACE sector level.}
	\label{fig:Distribution_of_clusters_int_macro}
\end{figure}

In Figure \ref{fig:Distribution_of_clusters_int_macro}, we plot the estimated distribution of the four clusters vulnerability scores. We observe that the extreme cluster, which includes sectors such as Mining and Quarrying, Electricity, Gas, Steam and Air Conditioning Supply, and Real Estate Activities, has considerably higher asset intensity --which implies higher exposure to climate risk-- than the other three.

\begin{table}[H]
	\scriptsize
	\centering 
	\begin{tabular}{l c c c}
		\toprule
		& \textbf{Available Companies} & \textbf{Average Intensity}&\textbf{Standard Deviation}\\
		\midrule
		\rowcolor{gray!15}
		Low & 1417 & 0.032 & 0.062\\
		Medium & 4733 & 0.071 & 0.104 \\
		\rowcolor{gray!15}
		High & 797 & 0.202 &  0.224 \\
		Extreme & 1476 & 0.345 &  0.338 \\
		\bottomrule
	\end{tabular}
	\\[10pt]
	\caption{\raggedright Descriptive analysis of asset intensity clustered in Low, Medium, High, and Extreme Risk.  We report the number of available companies, the average vulnerability score, and the standard deviation. Descriptive statistics are computed at the company level.}
	\label{table:Intensity_Table_Cluster}
\end{table}

\subsection{Expected Loss in a climate-stressed scenario}
\label{Shock}

In this section, we propose a simple methodology to compute the average loss under a climate-stressed scenario for the clusters defined in the previous section. 
We use this average loss to calibrate the jump term of the stressed firm value.

To obtain an estimation of the stressed equity values, we adopt the Gordon Growth Model \citep{Gordon56}, which establishes a connection between the equity value $E$ and the expected dividends in the subsequent period, assuming a constant growth rate in dividends.\\ In the \citet{Gordon56} model, the equity value is
\begin{equation}
	\label{eq:Gordon}
	E = \dfrac{{D}_0(1+g)}{q-g},
\end{equation}
where ${D}_0$ is the dividend at the initial time, $g$ is the growth rate of dividends, and $q$ is the required rate of return on the stock. We estimate the rate of return $q$  by obtaining the CAPM-$\beta$ \citep{sharpe1964capital} for each stock and multiplying it with the market return of the reference  index for the region. Moreover, we compute the growth rate of dividends $g$ as the slope of a regression line of EBITDA over time $t$
\begin{equation}
	\label{eq:EBITDA_slope}
	EBITDA_t = \beta_0 + g \cdot t + \epsilon_t,
\end{equation}
where $\epsilon_t$ is an error term.

In our climate-stressed scenario, consistently with \citet{bressan2024asset}, we assume that the physical risk impairs the firm's growth rate. The equity shocked price   $\Tilde{E}$ is obtained  from  equation \eqref{eq:Gordon} ,
 under the assumption that the rate of growth becomes $\Tilde{g}=(1-\alpha)g$, where $\alpha$ quantifies the climate shock for the stock.
Then, the shocks can be obtained as
\begin{equation}
	\label{eq:Shock}
	\text{shock} = \dfrac{\tilde{E}-E}{E} 
	= \dfrac{\tilde{E}}{E}-1 
	= \dfrac{1+\Tilde{g}}{q-\Tilde{g}} \cdot \dfrac{q-g}{1+g}-1\;\;,
\end{equation}
and
\begin{equation*}
	\label{eq:Shock_equity}
	\Tilde{E} = (1+\text{shock})E\;\;.
\end{equation*}
Notice that the shock value does not depend on the initial dividend ${D}_0$ but only on the growth rate of dividends $g$, the rate of return $q$, and the shock parameter $\alpha$. This stems from the use of the Gordon model to measure the impact of a penalized dividend growth rate on equity value, rather than to determine the equity value itself.

We consider a different $\alpha$ for each cluster that hereinafter we denote with the subscript $k$. We define $\alpha_k$ as the product of vulnerability and intensity score of the corresponding cluster $k$ (e.g. for the Mid-High vulnerability and Extreme asset intensity cluster $\alpha_k=0.345\cdot0.534=0.183$). Under this assumption, clusters with high vulnerability and high asset intensity are the most penalized. We report the value of $\alpha_k$ for each cluster in Table \ref{table:Alphas}.
\begin{table}[H]
	\scriptsize
	\centering 
	\begin{tabular}{l c c c c}
		\toprule
		& \textbf{Low} & \textbf{Medium}&\textbf{High}&\textbf{Extreme}\\
		\midrule
		\rowcolor{gray!15}
		Low & 0.006 & 0.018 & 0.042 & 0.073 \\
		Medium-High & 0.015 & 0.046 & 0.107 & 0.183 \\
		\bottomrule
	\end{tabular}
	\\[10pt]
	\caption[Shock Factors $\alpha$ in the Clusters]{\raggedright Shock factors $\alpha_k$ computed for each combination of vulnerability and intensity clusters. The rows correspond to vulnerability clusters, while the columns represent intensity clusters.}
	\label{table:Alphas}
\end{table}

\subsection{FVM estimation}
\label{sect:FVMEstimation}

In this section, we calibrate the FVM parameters in Equation \ref{eq:JumpMultiMertonSDE_Sol}. 
The calibration process is based on two steps.

First, we calibrate the model without the jump term $L_t$ directly on market data. 
The calibration follows the approach of  \citet{Moody2019}: we estimate the asset value $V_j(0)$ and the volatility $\hat{\sigma}_j$ of the FVM.
The approach is based on solving together equation \eqref{eq:BSFormula} and 
\begin{equation*}
        E_j(0)  \sigma_{E j} =\Phi(d_j)V_j(0) \hat{\sigma}_j \;\;,
\end{equation*}
where $d_j = \dfrac{log(V_j(0)/D_j)+(r_f+\frac{1}{2}\hat{\sigma}_j ^2)T_j}{\hat{\sigma}_j  \sqrt{T_j}}$, 
 $\sigma_{E_j}$ is the volatility of the equity  value $E_j$, $D_j$ is the total debt and $T_j$ the time to maturity. $V_j(0)$ and $\hat{\sigma}_j$ are obtained solving  numerically the non-linear system 
 \begin{equation*}
    \begin{aligned}
     E_j(0) &= C^{BS}(V_j(0), \hat{\sigma}_j,0,T_j,D_j) \\
     E_j(0)&=\frac{\Phi(d_j)V_j(0)\hat{\sigma}_j}{\sigma_{Ej}}  \;\;
    \end{aligned}
 \end{equation*}
for each stock $j$.
Secondly, for each climate cluster, we calibrate the intensity $\lambda_k$ of the Poisson process and the jump amplitude $\theta_k$. 
For each cluster, we select the parameters  $\lambda_k$ and $\theta_k$ that minimize the distance between the stressed equity value $\tilde{E}$ and the stressed-FVM for the stocks in the cluster.
Specifically, we minimize the root mean square percentage error (rMSPE)

\begin{equation*}
	\begin{aligned}
		& \argmin_{\lambda_k, \theta_k} \sqrt{ \sum_{j \in J_k} \dfrac{1}{\Tilde{E}_{j}^2(0)} \left[\Tilde{E}_{j}(0) - C_{LW}(\cdot, \lambda_k, \theta_k) \right]^2},
	\end{aligned}
\end{equation*}

where  $J_k$ is the set of companies belonging to cluster $k$, where for the sake of readability $C_{LW}(\cdot, \lambda_k, \theta_k):=C_{LW}(V_j(0),D_j,\hat{\sigma}_j,T_j,r_j,\theta_k,\lambda_k)$. We utilize the rMSPE because it accounts for the different orders of magnitude of different equity values.

In Table \ref{table:comp_pop}, we report a comparison between the climate-stressed expected loss for each cluster (reported in round brackets) with the one induced by the calibrated model. The values are expressed in absolute value as percentages. We observe that the average expected losses are close to the target.

\begin{table}[H]
	\scriptsize
	\centering 
	\begin{tabular}{l c c c c}
		\toprule
		& \textbf{Low} & \textbf{Medium}&\textbf{High}&\textbf{Extreme}\\
		\midrule
		\rowcolor{gray!15}
		Low & 1.51 & 3.75 & 6.62 & 11.57  \\ 
		\rowcolor{gray!15}
		& (1.69) & (4.42) & (7.76) & (13.00) \\ 
		Medium & 6.72 & 9.34 & 15.37 & 27.17  \\ 
		& (5.56) & (12.19) & (18.57) & (29.98) \\ 
		\bottomrule
	\end{tabular}
	\\[10pt]
	\caption[Comparison with Calibrated Shock - Mean (rMSPE)]{\raggedright Comparison between the climate stressed expected loss for each cluster (reported in round brackets) with the one induced by the calibrated model. The values are expressed in absolute value as percentages. The rows correspond to vulnerability clusters, while the columns represent intensity clusters.}
	\label{table:comp_pop}
\end{table}


We underline that, during the calibration procedure, we exclude those companies for which data are not fully available.
The initial dataset comprises 10,430 holding companies. Due to missing data we end up reducing the dataset to 5,351. In particular, we exclude  2,007 companies at  the climate clustering stage (section \ref{Cluster_vul}), 1,924 for the computation of the shocks (section \ref{Shock}), and 1,148 when estimating the Merton model (section \ref{sect:FVMEstimation}). In Table \ref{table:comp_pop2}, we report the distribution across clusters of the observations at the last two steps. We stress that we do not need to calibrate the Merton model for each company but only for the 8 clusters and we have at least 200 observations for each cluster.

\begin{table}[H]
	\scriptsize
	\centering 
	\begin{tabular}{l c c c c}
		\toprule
		& \textbf{Low} & \textbf{Medium}&\textbf{High}&\textbf{Extreme}\\
		\midrule
		\rowcolor{gray!15}
		Low & 490 & 1934 & 262 & 587  \\ 
		\rowcolor{gray!15}
		& (802) & (2602) & (404) & (810) \\
		Medium & 344 & 1240 & 209 & 285  \\ 
		& (494) & (2608) & (302) & (401) \\ 
		\bottomrule
	\end{tabular}
	\\[10pt]
	\caption[Observations by Cluster]{\raggedright Comparison between the available companies in each cluster at the end of climate clustering (reported in round brackets) with the one available for the FVM calibration. The rows correspond to vulnerability clusters, while the columns represent intensity clusters.}
	\label{table:comp_pop2}
\end{table}

\section{Evaluation of physical risk impacts}
\label{section:results}

In this section, we analyze the impact of climate risk on different equity portfolios.
We discuss the results for a set of representative indexes that cover global markets, as well as high- and low-exposed sectors toward climate risk.
Following our procedure, we can quantify an add-on for the VaR which can be directly added to the metric computed with the financial institution's internal models.




The methodology outlined in this paper, while utilized for VaR assessment, is designed to be adaptable and can be readily extended to the computation of Expected Shortfall (ES). This flexibility ensures compliance with evolving regulatory standards.

We compare the loss distributions without considering climate risk (baseline scenario) and under our climate-stressed scenario.
In the first case, i.e. the baseline Vasicek model in Equation \eqref{eq:MultiMertonSDE_Sol}, the percentage loss $L(t)$ at time $t$ for an index with weights $x_j$ is given by

\begin{equation}
	\label{eq:Perc_Ptf_loss_usual}
	L(t) = - \sum_{j \in J} x_j \left[ \dfrac{E_{j}(t)}{E_{j}(0)}-1 \right] \cdot 100 \;\;,
 \end{equation}

where $J$ is the set of all considered stocks. We underline that, consistently with market practices, when we evaluate the equity at time $t$, we assume that the corporate rolls its debt (i.e. the time-to-maturity $T_j$ is constant in time).
In the second case, we model the equity value considering the stressed model in \eqref{eq:JumpMultiMertonSDE_Sol}. We assume that the asset manager continues to evaluate the equity under the standard model in \eqref{eq:MultiMertonSDE_Sol}: i.e. she keeps neglecting physical risk in its internal model. Therefore, we evaluate the equity with the Gaussian model in \eqref{eq:MultiMertonSDE_Sol} while considering a stressed value for the company assets $\tilde{V}_j(t)$. The equity value at time $t$ in the stressed scenario is \[\tilde{E}_j(t)=C_{BS}(\tilde{V}_j(t),D_j,\hat{\sigma}_j,T_j,r_j)\;\;.
\]
The loss $\tilde{L}(t)$ at time $t$ under the stressed scenario is computed comparing  the  shocked equity at time $t$, $\tilde{E}_j(t)$ with the initial value $E_{j}(0)$
\begin{equation}
	\label{eq:Perc_Ptf_loss_jump}
	\tilde{L}(t) = - \sum_{j \in J} x_{j}\left[ \dfrac{\tilde{E}_{j}(t)}{E_{j}(0)}-1 \right] \cdot 100 \;\;.
\end{equation}

Notice that we compute the loss always with respect to the initial $E_{j}(0)$ to make consistent comparisons.

We  simulate portfolio losses in the two scenarios and compute the loss distributions.
We consider a full Monte Carlo approach \citep[see for reference][Section 2.3.3]{McNeil}, i.e. we compute the stock price at time $t$ for each simulation.
To compute the VaR, we run $10^5$ simulations for both the baseline and stressed models. We report in Algorithm \ref{alg:Computation_Loss}  the detailed steps of the computation of the equity value through the application of the Vasicek model.

In our framework, the dynamics of the asset values remain consistent with Vasicek’s risk-neutral paradigm, since the underlying stochastic evolution is not altered. The crucial difference, however, lies in debt management: instead of a single maturity, a continuous (rolling) refinancing mechanism is adopted, which continually extends the maturity. 
This approach better aligns with standard corporate practice.
Corporates commonly manage debt rollover through a process known as debt refinancing, which involves issuing new debt to repay maturing obligations. This practice allows firms to maintain liquidity, optimize capital structure, and potentially benefit from favorable interest rates \citep{valenzuela2016rollover}.
Consequently, this rolling structure leads to an equity payoff that differs from the classical formulation, where maturity is predetermined. As a result, while preserving internal consistency with the risk-neutral framework, the measure of the expected loss may differ from the one determined under the traditional Vasicek model. \footnote{We have also repeated the analysis under the assumption of fixed debt maturity. The results remain consistent with the baseline case, although we observe higher expected losses. This increase is attributable to the loss of the time-value component of the equity, which is modeled as a European call option on the firm’s asset value.
}


We analyze the simulation results for the $\Delta L = \mathbb{E}[\tilde{L}(t)]-\mathbb{E}[{L(t)}]$ and the ${\Delta VaR} = VaR \; \tilde{L}(t) - VaR\; L(t) \;$ at 90\%, 95\%, 99\% level respectively for different time-horizons $t=1,\,5,\,10,\,20$ years and for different indexes. The data concerning the composition of the indexes is based on information as of  31 December 2023.
In Table \ref{tab:msci_world}, we present the results for three relevant indexes: 
 the MSCI World index which represents a broad benchmark covering developed markets, serving as a standard reference for global equity performances; the MSCI World ESG Leaders and Climate Paris Aligned indexes which offer specialized versions, with the ESG Leaders index focusing on companies excelling in environmental, social, and governance practices, and the Climate Paris Aligned index prioritizing firms aligned with the Paris Agreement’s climate targets. 
The results are very similar because the three indexes exhibit comparable asset intensity and regional vulnerability profiles, as shown in Table \ref{tab:msci_clustering}.

At a 1-year horizon, all indexes show similar increases in $\Delta L$ and $\Delta VaR$, with VaR growing across different confidence levels. Over longer horizons, the differences become more pronounced, and both $\Delta L$ and $\Delta VaR$ increase over time for all indexes. Notably, at longer time horizons, such as 20 years, $\Delta VaR$ does not increase with higher quantiles. This is because, at extended time horizons, the equity value, which is modeled as a European Call option on the asset value, is very close to zero  for extreme quantiles of both processes with and without jumps. Therefore, there is no significant difference for extreme quantiles, while we still see an impact on the $\Delta L$.
We point out that, for an asset manager perspective, the relevant time-horizons is from one to five years for which the $\Delta VaR$ increases over quantiles.
 In this analysis, we assume no correlation between the L\'evy drivers $L_k(t)$ that model the climate shocks for each  cluster.  
For robustness check, in Appendix B, we focus on the extreme correlation case, i.e. when the correlation between the L\'evy drivers is the highest possible.
 We do not find significant differences in the case of correlated jump processes.

\begin{table}[]
\centering
\begin{tabular}{llllll}
\toprule
Index & Time horizon & $\Delta L$ & $\Delta VaR_{90\%}$ & $\Delta VaR_{95\%}$ & $\Delta VaR_{99\%}$ \\
\hline
 MSCI World & 1  & 0.61\%  & 0.84\%  & 1.08\%  & 2.49\%  \\
            & 5  & 3.23\%  & 3.30\%  & 3.63\%  & 4.50\%  \\
            & 10 & 7.55\%  & 5.46\%  & 5.85\%  & 6.52\%  \\
            & 20 & 20.97\% & 11.66\% & 11.13\% & 9.52\%  \\
\hline
 MSCI World ESG Leaders & 1  & 0.57\%  & 0.82\%  & 1.04\%  & 2.51\%  \\
                        & 5  & 3.02\%  & 3.09\%  & 3.54\%  & 4.15\%  \\
                        & 10 & 7.06\%  & 5.49\%  & 5.70\%  & 5.97\%  \\
                        & 20 & 18.61\% & 10.68\% & 10.13\% & 8.50\%  \\
\hline
 MSCI World Climate Paris Aligned & 1  & 0.63\%  & 0.85\%  & 1.09\%  & 2.58\%  \\
                                  & 5  & 3.32\%  & 3.40\%  & 3.48\%  & 4.27\%  \\
                                  & 10 & 7.68\%  & 5.49\%  & 5.74\%  & 5.90\%  \\
                                  & 20 & 20.06\% & 10.99\% & 10.05\% & 8.16\%  \\
\bottomrule
\end{tabular}
\\[10pt]
\caption{\raggedright $\Delta L$, ${\Delta VaR_{90\%}}$, ${\Delta VaR_{95\%}}$, and ${\Delta VaR_{99\%}}$ for MSCI World (A), MSCI World ESG Leaders, and MSCI World Climate Paris Aligned indexes over various time horizons. Results are derived from simulated dynamics under the risk-neutral measure.}
\label{tab:msci_world}
\end{table}
\begin{table}[]
\centering
\renewcommand{\arraystretch}{1.2}

\begin{tabular}{lcccc}
\toprule
A & {Low} & {Medium} & {High} & {Extreme} \\
\midrule
Low & 9.06\% & 72.61\% & 3.31\% & 7.84\% \\
Medium-High & 0.79\% & 5.53\% & 0.35\% & 0.51\% \\
\bottomrule
\end{tabular}

\vspace{10pt}

\begin{tabular}{lcccc}
\toprule
B & {Low} & {Medium} & {High} & {Extreme} \\
\midrule
Low & 7.56\% & 74.73\% & 4.38\% & 5.71\% \\
Medium-High & 1.15\% & 5.79\% & 0.30\% & 0.38\% \\
\bottomrule
\end{tabular}

\vspace{10pt}

\begin{tabular}{lcccc}
\toprule
C & {Low} & {Medium} & {High} & {Extreme} \\
\midrule
Low & 8.94\% & 74.57\% & 1.09\% & 7.75\% \\
Medium-High & 0.53\% & 4.45\% & 2.04\% & 0.62\% \\
\bottomrule
\end{tabular}
\\[10pt]

\caption{\raggedright Clustering results for MSCI World (A), MSCI World ESG Leaders (B), and MSCI World Paris Aligned (C) respectively. The rows correspond to vulnerability clusters, while the columns represent intensity clusters.}
\label{tab:msci_clustering}
\end{table}

\begin{figure}[]
	\centering
	\includegraphics[scale=0.8]{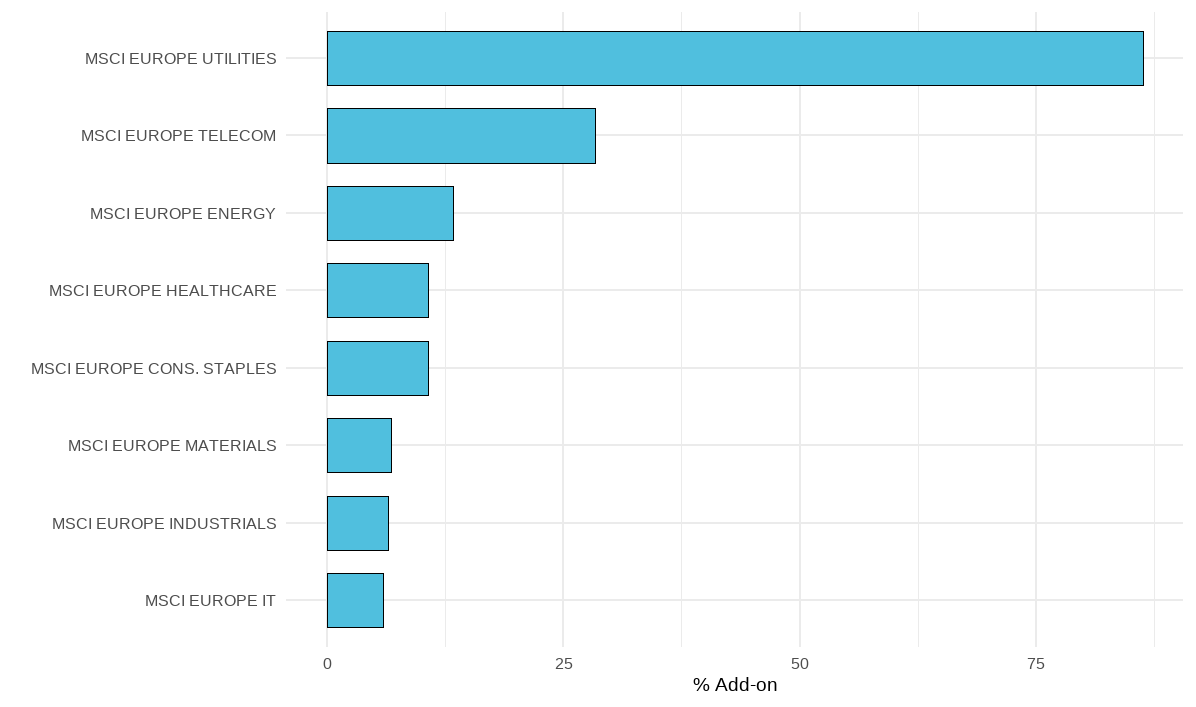}
	\caption[Comparison of \% Add-on on VaR 95\% on 5 years]{\raggedright Comparison of \% Add-on on 5 year VaR 95\%.}
	\label{fig:comp_add_on}
\end{figure}

We now focus on comparing the add-ons between different European sectoral indexes. These indexes, part of the MSCI Europe family, represent the performance of various industries within developed European markets. By selecting European indexes, we eliminate the effect of climate vulnerability specific to geographic regions, isolating the intrinsic asset intensity of each sector. This approach enables a more targeted evaluation of sector-specific exposures to physical climate risks, offering a clearer view of the mostly affected sectorial portfolios. 
To do so, we consider only the confidence interval at $95\%$ at 5 years and compare the percentage add-on computed as
\begin{equation*}
	\% \ \text{Add-on (t)}=\left( \frac{{VaR \;\tilde{L}(t)}}{VaR \; L(t)}-1 \right) \cdot 100\;\;.
\end{equation*}
In Figure \ref{fig:comp_add_on}, we present these add-on terms ordered from highest to lowest.

\begin{figure}[]
	\centering
	\includegraphics[scale=1]{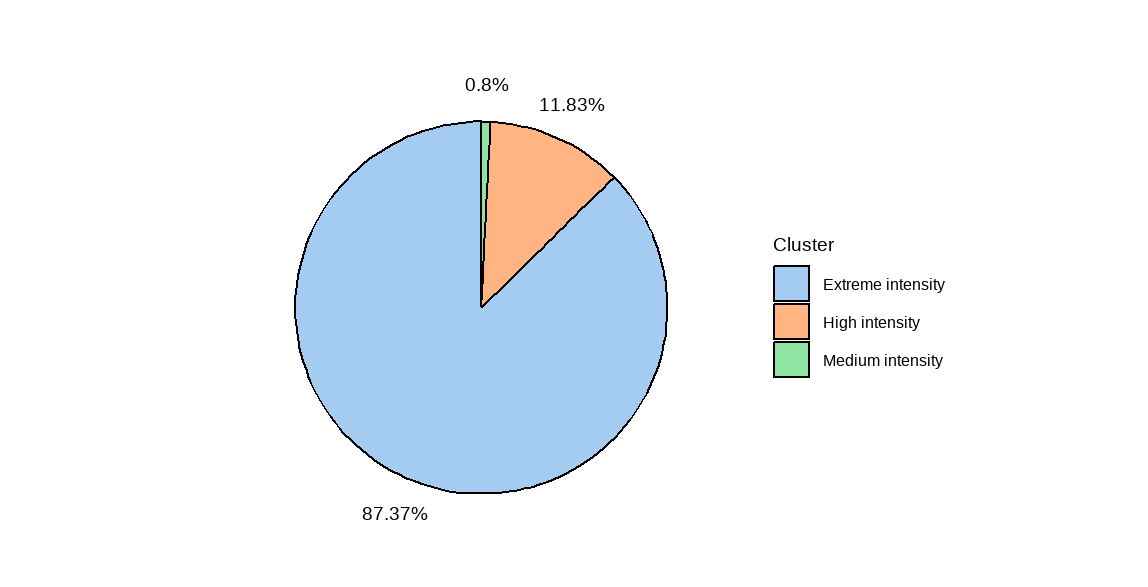}
	\caption[Composition of MSRLUTI index]{\raggedright MSCI Europe Utilities Index portfolio composition in terms of the asset intensity cluster. Note that all the stocks in this index belong exclusively to the low-vulnerability geographical cluster.}
	\label{fig:pie_MSRLUTI}
\end{figure}

The utilities sector stands out as the most exposed, with the highest add-on of 86\%. This highlights its substantial sensitivity to physical climate risks, such as extreme weather events, which can cause significant disruptions to critical infrastructure. This conclusion is further supported by the cluster analysis of asset intensity (Figure \ref{fig:pie_MSRLUTI}), which reveals that the majority of firms in the utility index is categorized under extreme intensity.\footnote{These findings are in line with a report by S\&P, which underscores that the utilities sector faces the greatest threat as climate risks intensify, see \url{https://www.spglobal.com/market-intelligence/en/news-insights/articles/2021/9/utilities-face-greatest-threat-as-climate-risks-intensify-66613890}.}

\begin{figure}[]
	\centering
	\includegraphics[scale=1]{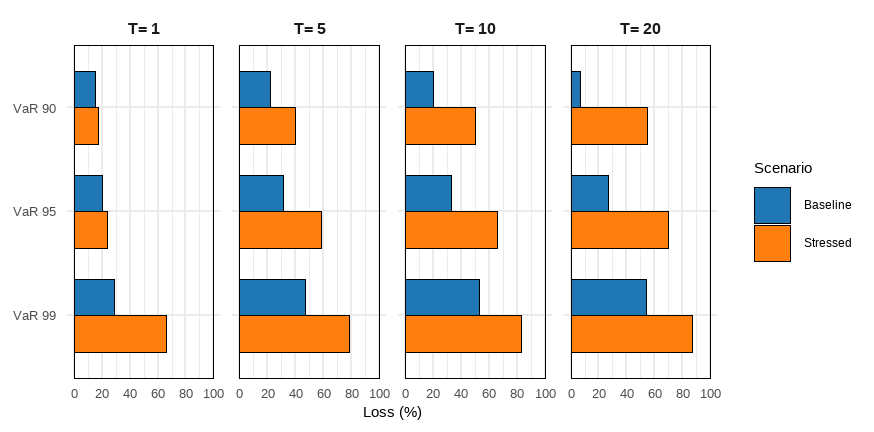}
	\caption[Comparison over Time of VaR for MSCI Europe Utilities index]{\raggedright Comparison over Time of VaR for the MSCI Europe Utilities index. We present comparison of VaR at confidence intervals of 90\%, 95\%, and 99\% over the different time horizons.  Blue bars represent the loss associated with the dynamic without jumps, while orange bars represent the loss accounting for them.}
    \label{fig:MSRLUTI}
\end{figure}

Figure \ref{fig:MSRLUTI} illustrates the $VaR$ values over the time horizons $t = 1, 5, 10, 20$ years. It is noteworthy that, over time, the $VaR$ of the stressed model continues to exhibit a negative trend, increasing its risk exposure as the years progress. In contrast, the baseline model shows a decreasing $VaR$ over the same time horizons, reflecting the positive long-term trend of equity growth under normal market conditions.




The telecommunications and energy sectors show add-ons of 28\% and 13\%, respectively. These sectors and infrastructures are also particularly susceptible to climate events due to their physical exposure and the need to operate under variable environmental conditions. For instance, storms can disrupt power lines and damage equipment required for data transmission, while floods can compromise data centers and telecommunication stations \citep{montoya2023socio}.

In Appendix C, we explain in detail the algorithm  for simulating the climate-stressed scenarios and computing  losses on various portfolios.
Moreover, we report additional robustness checks regarding the computation of losses when the investor does not neglect climate risk. 



\newpage
\section{Conclusions \& Policy implications}
\label{conclusion}
In this paper, we introduce a novel and easily implementable model framework for quantifying the impact of physical climate risk on an equity portfolio, utilizing an FVM approach.
 The framework is designed to be seamlessly integrated into the decision-making processes of financial institutions and can be tailored to support a variety of climate risk metrics, depending on the specific regulatory or internal risk management objectives.

Our approach extends the multivariate Vasicek model in \eqref{eq:MultiMertonSDE_Sol} by incorporating negative jumps to capture the sudden and severe financial impacts associated with extreme climate events. An explicit formula for the stressed equity value, based on the characteristic function of the jump process, is provided in \eqref{eq:option price}. 
 These negative jumps are introduced to model abrupt losses in firm value that may result from physical damages or operational disruptions. For instance, a wildfire might destroy a firm's warehouse, leading to immediate asset write-downs, while an electricity shortage caused by a heatwave could halt production and severely impair revenue streams. By explicitly accounting for such climate-induced jumps, the proposed framework enhances the realism and robustness of a traditional market risk model, making it more suitable for analyzing portfolio physical risk. We underline that the model is constructed such that climate-related losses directly impact the firm's asset value, thereby ensuring that the transmission channel to equity prices is explicitly and transparently reproduced. This direct linkage allows for a more accurate assessment of how extreme physical climate events propagate through to market valuations.

We calibrate the model using estimated climate risk losses in conjunction with widely available balance sheet data, such as total assets and asset intensity. To account for firm-level heterogeneity in exposure to physical climate risk, we classify companies into clusters based on two key dimensions: geographical vulnerability and asset intensity. Specifically, we construct two vulnerability clusters—distinguishing firms operating in high- versus low-risk regions—and four clusters representing increasing levels of asset intensity (cf. Figures~\ref{fig:Distribution_of_clusters}–\ref{fig:Distribution_of_clusters_int_macro} and Tables~\ref{table:Vulnerability_Table_Cluster_2}–\ref{table:Intensity_Table_Cluster}). We highlight that the cluster associated with high vulnerability scores corresponds almost perfectly to countries in the Global South. This empirical finding reflects the well-documented asymmetry in the geographic distribution of climate-related physical risks, with developing economies being disproportionately affected by extreme weather events, rising temperatures, and sea-level rise.

The model parameters are then estimated for each cluster using the two-step procedure described in Section~\ref{sect:FVMEstimation}, allowing for cluster-specific dynamics in response to climate shocks. The decision to group firms by both asset intensity and geographical exposure is motivated by the expectation that companies located in similar regions and with similar shares of physical assets will be affected in comparable ways by extreme climate events. This clustering strategy enhances the model's ability to capture intra- and inter-sectoral dependencies, which is essential for producing more realistic portfolio-level risk assessments.

We apply the calibrated model to compute the differences in Value-at-Risk (VaR) and expected loss between a baseline scenario and a climate-stressed scenario across various portfolios (see Table \ref{table:Alphas} and Figure \ref{fig:MSRLUTI}). 
We analyze simulation results for expected losses ($\Delta L$) and changes in Value-at-Risk ($\Delta VaR$) at the 90\%, 95\%, and 99\% confidence levels over time horizons of 1, 5, 10, and 20 years, across three major indexes: MSCI World, MSCI World ESG Leaders, and MSCI Climate Paris Aligned. These indexes exhibit similar asset intensity and regional vulnerability profiles, resulting in comparable climate risk impacts.
At short horizons (e.g., 1 year), $\Delta L$ and $\Delta VaR$ show consistent increases across all indexes and quantiles. No significant differences were observed between the MSCI World index and its ESG-aligned versions, highlighting that investing in ESG assets does not necessarily offer protection against the financial impacts of physical climate risk.\footnote{For a review of ESG scores see e.g. \cite{ferro2025uncovering}.}
Building on the VaR computation, we propose an add-on to the portfolio VaR that explicitly accounts for physical climate risk (see Figure \ref{fig:comp_add_on}). This framework highlights the heterogeneous exposure to climate risk across sectors such as utilities, which exhibit higher vulnerability, with the highest VaR add-on of 86\%. Our finding is motivated analyzing  the firms that are included in the MSCI utility index  (Figure~\ref{fig:pie_MSRLUTI}). We observe that the majority of firms within the index fall into the extreme intensity category. These firms typically maintain large, immobile physical assets—such as power plants, transmission networks, and water systems—which are particularly vulnerable to climate-induced disruptions.

From a policy perspective, our framework enables the computation of a Value-at-Risk (VaR) add-on for any equity portfolio, provided that the geographical location, sector classification, and widely available balance sheet data of the constituent stocks are known. This facilitates the definition of standardized VaR add-ons by regulatory authorities, which could potentially be incorporated into the Basel standardized approach. Importantly, this would not require supervised entities to independently implement or replicate the underlying climate risk model. 
As an illustrative example, regulators could compute VaR add-ons using representative portfolios composed of the most asset-intensive stocks available in a given market (e.g., the U.S. equity market). The resulting add-ons could serve as conservative benchmarks, applicable to all portfolios that are exclusively exposed to the same geographical region.
Moreover, we highlight that our calibration approach facilitates regulatory and policy applications, as it enables authorities to derive generalized insights on climate vulnerability patterns and apply them to supervisory stress testing exercises. It also reduces model complexity for institutions by allowing them to apply pre-calibrated cluster-level parameters based on simple, observable firm characteristics.


In summary, these results address the practical needs of financial institutions and respond to regulatory expectations—such as those set forth in the Basel framework—to effectively incorporate physical climate risks into risk management practices.

\subsection*{Limitations and Further developments}

While the proposed framework provides a tractable and flexible approach to modelling physical climate  market risk, it also presents several limitations that open avenues for future research. A primary constraint lies in the geographical treatment of asset exposure: we consider only the location of firms' headquarters, neglecting the precise geographical distribution of individual assets. This simplification may lead to misclassification of climate exposure, especially for multinational firms with physical assets spread across diverse regions. A natural extension, subject to data availability, would be to assign assets to different climate clusters based on their specific geographical positions or to determine a firm's location based on where the majority of its physical assets are located.

Another modelling limitation stems from the assumption that climate shocks follow a L\'evy process. While L\'evy processes offer analytical tractability and encompass a broad class of jump processes, they are characterized by independent and stationary increments. Such assumptions may be too simplistic for modelling climate phenomena, which tend to be non-stationary and exhibit temporal dependencies. To better capture these features, the framework could be extended by incorporating {additive processes} - processes with independent but non-stationary increments (see, e.g., \citet{azzone2022additive}) - or {Hawkes processes} - self-exciting processes in which past events increase the likelihood of future shocks (see, e.g., \citet{hawkes2018hawkes}). These alternatives could provide a more realistic representation of the complex, path-dependent nature of climate dynamics.

\bibliographystyle{elsarticle-harv} 
\bibliography{sources}
\bigskip

\section*{Acknowledgments}
We thank N. Bartolini, R. Baviera, G. Bressan, R. Ceretti, U. Cherubini, F. Maglione,  P. Manzoni, E. Sala, A. Sbuelz, L. Taschini, T. Vargiolu for the fruitful conversations on the topic and all the participants to the III International Fintech Conference 2025, the Energy Finance X Conference and the CLIFIRIUM 2025 Conference. 
Michele Azzone and Davide Stocco are members of the Gruppo Nazionale Calcolo Scientifico-Istituto Nazionale di Alta Matematica (GNCS-INdAM).
The research of Davide Stocco was carried out in the framework of ``Energy for Climate" interdisciplinary research center and was supported financially by the ``Decarbonize energy" program of the Institut Polytechnique de Paris.
\newpage

\section*{Appendix A: Dataset descriptive statistics }

In Table \ref{tab:apxA1}, we report a snapshot of 16 firms in the dataset. 
For each of the 16 randomly extracted firms we report the ISIN code, the Country, the NACE sector code, the MSCI index they are part of, the Vulnerability and the Intensity cluster the total debt (in billion local currency) the TTM (in years) and the market capitalization (in billion local currency).

\begin{table}[h]\resizebox{\columnwidth}{!}{%
\begin{tabular}{lllllllll}
\toprule
\textbf{ISIN}         & \multicolumn{1}{c}{\textbf{Country}} & \multicolumn{1}{c}{\textbf{NACE}} & \multicolumn{1}{c}{\textbf{Index}} & \multicolumn{1}{c}{\textbf{ Vulnerability}} & \multicolumn{1}{c}{\textbf{Intensity}} & \multicolumn{1}{c}{\textbf{Debt}} & \multicolumn{1}{c}{\textbf{TTM}} & \multicolumn{1}{c}{\textbf{Mkt Cap}} \\
\hline
CA82509L1076 & Canada                               & G                                 & MSCI USA                          & Low                                                  & Low                                            & 0.9                               & 2.5                              & 99.1                                 \\
FR0000039299 & France                               & M                                 & MSCI Europe                       & Low                                                  & Low                                            & 7.8                               & 4.5                              & 16.5                                 \\
PHY806761029 & Philippines                          & G                                 & MSCI Emerging Asia                & Medium-High                                               & Low                                            & 541.1                             & 3.8                              & 1050.4                               \\
SA15GG53GHH3 & Saudi Arabia                         & M                                 & MSCI EMEA                         & Medium                                               & Low                                            & 0.1                               & 6.4                              & 63.3                                 \\
US7443201022 & USA             & K                                 & MSCI USA                          & Low                                                  & Medium                                         & 21.1                              & 8.9                              & 38.2                                 \\
US1729674242 & USA             & K                                 & MSCI USA                          & Low                                                  & Medium                                         & 501.8                             & 5.1                              & 99.6                                 \\
BRGGBRACNPR8 & Brazil                               & C                                 & MSCI Emerging America             & Medium-High                                               & Medium                                         & 13.6                              & 9.1                              & 27.3                                 \\
JP3818000006 & Japan                                & J                                 & MSCI Pacific                      & Medium-High                                               & Medium                                         & 211.2                             & 0.5                              & 4142.4                               \\
US9078181081 & USA             & H                                 & MSCI USA                          & Low                                                  & High                                           & 33.3                              & 14.6                             & 151                                  \\
GRS419003009 & Greece                               & R                                 & MSCI EMEA                         & Low                                                  & High                                           & 0.8                               & 6.4                              & 5.4                                  \\
KYG4712E1035 & China                                & Q                                 & MSCI Emerging Asia                & Medium-High                                               & High                                           & 1.3                               & 3.8                              & 21.8                                 \\
BRRDORACNOR8 & Brazil                               & Q                                 & MSCI Emerging America             & Medium-High                                               & High                                           & 37.2                              & 8.2                              & 64.8                                 \\
AN8068571086 & USA             & B                                 & MSCI USA                          & Low                                                  & Extreme                                        & 12.2                              & 8                                & 73.8                                 \\
US5529531015 & USA             & I.55                              & MSCI USA                          & Low                                                  & Extreme                                        & 8.9                               & 3.7                              & 17.2                                 \\
SG1U68934629 & Singapore                            & D                                 & MSCI Pacific                      & Medium-High                                               & Extreme                                        & 10.4                              & 5.5                              & 12.4                                 \\
CNE000000T18 & China                                & B                                 & MSCI Emerging Asia                & Medium-High                                               & Extreme                                        & 7.8                               & 3.8                              & 69.9     \\
\bottomrule
\end{tabular}}\caption{Snapshot of firms in the dataset. For the 16 randomly extracted firms we report the ISIN code, the Country, the NACE sector code, the MSCI index they are part of, the Vulnerability and the Intensity cluster the total debt (in billion local currency) the TTM (in years) and the market capitalization (in billion local currency). }
\label{tab:apxA1}
\end{table}

In Table \ref{tab:apxA2} we report descriptive statistics (mean, standard deviation, median, 5th and 95th percentiles) of Asset Intensity and Geographical Vulnerability computed at the country level. The reported moments are therefore obtained by averaging across countries. In addition, we provide descriptive statistics for the number of firms per country and for the number of firms per sector, to highlight the distribution of firms in our dataset across geographical and sectoral dimensions.

\begin{table}[h]
\begin{tabular}{llllll}
\toprule
                     & Mean     & Standard Deviation & Median   & Q0.05   & Q0.95    \\
                     \hline
Asset Intensity per country     & 1.42 & 3.29         & 0.82& 0.09& 3.89 \\
Vulnerability  per country      & 0.44  & 0.23            & 0.45 & 0.11 & 0.92  \\
\# Firms per country & 148.99 & 344.66           & 14.50     & 1        & 521.40    \\
\# Firms per sector  & 16.26 & 36.21           & 6        & 1        & 62.05   \\
\bottomrule
\end{tabular}\caption{We report descriptive statistics (mean, standard deviation, median, 5th and 95th percentiles) of Asset Intensity and Geographical Vulnerability computed at the country level. The reported moments are therefore obtained by averaging across countries. In addition, we provide descriptive statistics for the number of firms per country and for the number of firms per sector, to highlight the distribution of firms in our dataset across geographical and sectoral dimensions. \label{tab:apxA2}}
\end{table}

\section*{Appendix B: Correlation Robustness}
\label{apxA}
In this Appendix, we discuss whether the correlation between the different L\'evy processes that model the climate shocks have a significant impact  on the final results.

Throughout the paper, we assume no correlation between the L\'evy drivers $L_k(t)$ that model the climate shocks.  
In the following, we focus on the other extreme case, i.e. when the correlation between the L\'evy drivers is the highest possible.
Without loss of generality, we consider the Poisson processes $N_k(t)$ sorted by ascending intensity, i.e. $\lambda_1\leq \lambda_2<...<\lambda_k<...\lambda_K$, where $K$ is the number of clusters considered. We set recursively
\begin{align*}
    N_{1}(t)&=\hat{N}_1(\lambda_1, t),\\
    \dots,&\\
    N_k(t)&= N_{(k-1)}(t)+\hat{N}_k(\lambda_k-\lambda_{k-1}, t)\;\;,
\end{align*}
where $\hat{N}_k(\lambda, t)$ is a Poisson process with parameter $\lambda$. Let us notice that it is possible to prove by induction that $N_{(k-1)}(t)+\hat{N}_k(\lambda_k-\lambda_{k-1}, t)$ is a Poisson process with parameter $\lambda_k$.
This result arises from the well-established property that the sum of two independent Poisson processes is itself a Poisson process, with an intensity equal to the sum of the individual intensities.

In Table \ref{tab:msci_world_jump_correlated}, we report the $\Delta L$ and $\Delta \text{VaR}_{90\%}$ in case of independent and correlated jumps for MSCI World, MSCI World ESG Leaders, and MSCI World Climate Paris Aligned indexes over the time horizons 1, 5, 10 and 20 years. Let us notice that considering correlation  $\Delta L$ and $\Delta \text{VaR}_{90\%}$ increases. However, results do not differ substantially (no more than 2\% in terms of  $\Delta \text{VaR}_{90\%}$, even in the 20 years time horizon).
Our results are therefore robust to the presence of correlation among different clusters.

\begin{table}[]
\centering
\begin{tabular}{lccccc}
\toprule
\multirow{2}{*}{}  & \multirow{2}{*}{} & \multicolumn{2}{c}{Independent Jumps} & \multicolumn{2}{c}{ Correlated Jumps} \\ Index
                  &   Time horizon   & $\Delta L$ & $\Delta \text{VaR}_{90\%}$ & $\Delta L_{}$ & $\Delta \text{VaR}_{90\%}$ \\
\hline
MSCI World & 1  & 0.61\%  & 0.84\%  & 0.61\%  & 0.86\%  \\
           & 5  & 3.23\%  & 3.30\%  & 3.21\%  & 3.31\%  \\
           & 10 & 7.55\%  & 5.46\%  & 7.47\%  & 5.86\%  \\
           & 20 & 20.97\% & 11.66\% & 20.28\% & 12.84\% \\
\hline
MSCI World ESG Leaders & 1  & 0.57\%  & 0.82\%  & 0.58\%  & 0.81\%  \\
                       & 5  & 3.02\%  & 3.09\%  & 3.01\%  & 3.05\%  \\
                       & 10 & 7.06\%  & 5.49\%  & 6.93\%  & 5.87\%  \\
                       & 20 & 18.61\% & 10.68\% & 18.36\% & 11.65\% \\
\hline
MSCI World Climate Paris Aligned & 1  & 0.63\%  & 0.85\%  & 0.63\%  & 0.87\%  \\
                                 & 5  & 3.32\%  & 3.40\%  & 3.28\%  & 3.45\%  \\
                                 & 10 & 7.68\%  & 5.49\%  & 7.54\%  & 6.11\%  \\
                                 & 20 & 20.06\% & 10.99\% & 19.88\% & 12.63\% \\
\bottomrule
\end{tabular}
\\[10pt]
\caption{\raggedright $\Delta L$ and $\Delta \text{VaR}_{90\%}$ in case of independent and correlated jumps for MSCI World, MSCI World ESG Leaders, and MSCI World Climate Paris Aligned indexes over the time horizons 1, 5, 10 and 20 years.}
\label{tab:msci_world_jump_correlated}
\end{table}


\newpage
\section*{Appendix C: Climate Stress Scenario Simulation}

In this Appendix, we explain in detail the algorithm  for simulating the climate-stressed scenarios and computing  losses on various portfolios.
Moreover, we report additional robustness checks regarding the computation of losses when the investor does not neglect climate risk. 

In Algorithm \ref{alg:Simulation_Jump_Asset}, we report the steps for simulating the asset value in \eqref{eq:JumpMultiMertonSDE_Sol} and \eqref{eq:MultiMertonSDE_Sol}, with and without jumps, respectively.
For each time step $dt$, we simulate  $N_{ptf}$ independent Brownian Motion $W_j$, one market component $Z$ and $K$ compound Poisson processes, where $N_{ptf}$ is the number of portfolios and $K$ is the number of clusters in the analysis.

\begin{algorithm}[]
	\caption{Simulation of the dynamic of the total asset with and without jumps}
	\label{alg:Simulation_Jump_Asset}
	\begin{algorithmic}[1]
		
		\State Initialize matrix of returns $X=X_{jump}=zeros(N_{ptf}, M+1)$
		\State Define time step $dt = T/M$
		\State{Simulate market component $Z\sim \mathcal{N}(0, 1)$ as vector of length M}
		\State{Simulate idiosyncratic component $W\sim \mathcal{N}(0, 1)$ as matrix of size $(N_{ptf}, M)$}
		\State{Simulate diffusion path
			$Diffusion = \left( r_f - \dfrac{\sigma^2}{2} - \dfrac{\omega^2}{2} \right) dt
			+\sigma\sqrt{dt}W
			+\omega\sqrt{dt}Z$}
		\State $X(:,2:end)=cumsum(Diffusion, 2)$
		\State $JumpDiffusion = Diffusion$
		\State Find the clusters present in the portfolio
		\For{over the clusters}
		\State{Simulate $N_t\sim Poisson(T \cdot \lambda)$}
		\State Define jumping times {$T_{jump}=sort(T \cdot rand(N_t,1))$}
		\For{over the resampling times}
		\For{over jumping times $T_{jump}$}
		\If{the process jumps at this resampling time}
		\State{Simulate the jump $Y$ and add it to JumpDiffusion}
		\EndIf                    
		\EndFor
		\EndFor
		\State $X^{jump}(rows,2:end)=cumsum(JumpDiffusion(rows,:), 2)$
		\EndFor
		\State Compute asset dynamic $V=V_0e^X$,  $V^{jump}=V_0e^{X^{jump}}$
	\end{algorithmic}
\end{algorithm}

Once we have simulated the value process over time in the baseline and stressed scenario, we compute the firm value under the assumption that the investor neglects climate risk.
In Algorithm \ref{alg:Computation_Loss}, we summarise the steps for computing the loss with and without jumps starting from the asset dynamic simulated as described in Algorithm \ref{alg:Simulation_Jump_Asset}.

\begin{algorithm}[]
	\caption{Computation of the loss with and without jumps}
	\label{alg:Computation_Loss}
	\begin{algorithmic}[1]
		
		\State Initialize $Loss$ and $Loss_{shock}$
		\For{over the number of simulations}
		\State Simulate asset dynamics $V$, $V^{jump}$ (see Algorithm~\ref{alg:Simulation_Jump_Asset})
		\For{over the asset in the portfolio}
		\State $E_T = Call^{BS}(V_T, Debt \;\; Maturity)$
		\State $E_T^{jump} = Call^{BS}(V_T^{jump}, Debt \;\; Maturity)$
		\EndFor
        \State Initialize $E_0$ calibrated with the FVM 
		\State $Loss = -\sum_{assests} weight \cdot \left[ \dfrac{E_T}{E_0}-1 \right] \cdot 100$
		\State $Loss_{shock} = -\sum_{assests} weight \cdot \left[ \dfrac{E_T^{jump}}{E_0}-1 \right] \cdot 100$
		\EndFor
	\end{algorithmic}
\end{algorithm}

Finally, we address the methodology employed to price the equity value under a stressed scenario.
In the analysis presented in section \ref{section:results}, 
we assume that the investor continues to evaluate firms based on their risk-neutral dynamics neglecting future climate events.
The results obtained using Black and Scholes reflect the perspective of an observer who remains unaware of climate risks and continues to employ traditional pricing techniques.

Hereinafter, we repeat the analysis by considering that the investor
incorporates physical risk at the evaluation date and thus prices the equity under the climate-stressed  dynamic for the asset value in \eqref{eq:JumpMultiMertonSDE_Sol}. 
This methodology reflects the perspective of an observer who acknowledges the presence of climate risks and incorporates them into the valuation process.

In Table \ref{tab:lewis_pricing}, we report the results for the $\Delta L$, $VaR_{90\%}$, $VaR_{95\%}$ and $VaR_{99\%}$ values calculated using the Lewis pricing formula for various MSCI indexes over 1, 5, 10 and 20 years.
We compare these results with the ones in Table \ref{tab:msci_world}.
As expected, the $\Delta VaR$ values obtained under the Lewis formula are systematically more pessimistic compared to those derived using Black and Scholes. This difference underscores the importance of incorporating climate risks into the pricing framework, as it provides a more realistic and prudent estimate of potential losses. 
However, let us point out that the differences are stronger for shorter maturities (up to 300\% higher than those in Table \ref{tab:msci_world}), while results do not differ significantly for time horizons longer than 5 years.

\begin{table}[h]
\small
\centering
\begin{tabular}{llllll}
\toprule
Index & Time horizon & $\Delta L$ & $\Delta VaR_{90\%}$ & $\Delta VaR_{95\%}$ & $\Delta VaR_{99\%}$ \\
\midrule
MSCI WORLD & 1  & 4.60\%  & 3.88\%  & 3.89\%  & 5.00\% \\
           & 5  & 7.52\%  & 5.52\%  & 5.65\%  & 5.78\% \\
           & 10 & 12.22\% & 7.67\%  & 7.49\%  & 7.19\% \\
           & 20 & 25.38\% & 12.65\% & 11.68\% & 9.52\% \\
\midrule
MSCI WORLD ESG LEADERS & 1  & 4.88\%  & 4.08\%  & 4.10\%  & 5.09\% \\
                       & 5  & 8.01\%  & 5.88\%  & 5.86\%  & 6.15\% \\
                       & 10 & 13.06\% & 7.75\%  & 7.73\%  & 7.79\% \\
                       & 20 & 28.30\% & 13.81\% & 12.82\% & 10.57\% \\
\midrule
MSCI WORLD CLIMATE PARIS ALIGNED & 1  & 5.16\%  & 4.29\%  & 4.29\%  & 5.33\% \\
                                 & 5  & 8.32\%  & 6.07\%  & 5.81\%  & 6.04\% \\
                                 & 10 & 13.36\% & 7.84\%  & 7.63\%  & 7.21\% \\
                                 & 20 & 27.37\% & 13.14\% & 11.67\% & 9.17\% \\
\bottomrule
\end{tabular}
\caption{\raggedright $\Delta L$, $VaR_{90\%}$, $VaR_{95\%}$ and $VaR_{99\%}$ values calculated using the Lewis pricing formula for various MSCI indexes over 1, 5, 10 and 20 years.}
\label{tab:lewis_pricing}
\end{table}

\end{document}